# Challenges for density functional theory in simulating metal-metal singlet bonding: a case study of dimerized VO$_2$

Yubo Zhang[1,*], Da Ke[1], Junxiong Wu[1], Chutong Zhang[1], Baichen Lin[2,3], Zuhuang Chen[4,5], John P. Perdew[6], Jianwei Sun[6,*]

[1]*Minjiang Collaborative Center for Theoretical Physics, College of Physics and Electronic Information Engineering, Minjiang University, Fuzhou, China*

[2]*School of Materials Science and Engineering, Nanyang Technological University, Singapore, 639798, Republic of Singapore*

[3]*Institute of Materials Research and Engineering (IMRE), Agency for Science, Technology and Research (A\*STAR), Singapore 138634, Republic of Singapore*

[4]*School of Materials Science and Engineering, Harbin Institute of Technology, Shenzhen, Shenzhen 518055, China*

[5]*Flexible Printed Electronics Technology Center, Harbin Institute of Technology, Shenzhen, Shenzhen 518055, China*

[6]*Department of Physics and Engineering Physics, Tulane University, New Orleans, Louisiana 70118, USA*

Corresponding emails: yubo.drzhang@mju.edu.cn, jsun@tulane.edu

VO$_2$ is renowned for its electric transition from an insulating monoclinic (M$_1$) phase characterized by V-V dimerized structures, to a metallic rutile (R) phase above 340 Kelvin. This transition is accompanied by a magnetic change: the M$_1$ phase exhibits a non-magnetic spin-singlet state, while the R phase exhibits a state with local magnetic moments. Simultaneous simulation of the structural, electric, and magnetic properties of this compound is of fundamental importance, but the M$_1$ phase alone has posed a significant challenge to density functional theory (DFT). In this study, we show none of the commonly used DFT functionals, including those combined with on-site Hubbard $U$ to better treat 3$d$ electrons, can accurately predict the V-V dimer length. The spin-restricted method tends to overestimate the strength of the V-V bonds, resulting in a small V-V bond length. Conversely, the spin-symmetry-breaking method exhibits the opposite trends. Each bond-calculation method underscores one of the two contentious mechanisms, i.e., Peierls or Mott, involved in the metal-insulator transition in VO$_2$. To elucidate the challenges encountered in DFT, we also employ an effective Hamiltonian that integrates one-dimensional magnetic sites, thereby revealing the inherent difficulties linked with the DFT computations.

## 1. Introduction

In transition-metal compounds, the coupling between the lattice-charge-spin-orbital degree of freedom makes these materials a fascinating playground for developing multiple functionalities [1]. At the microscopic level, these interesting properties originate from or are associated with correlated electrons in the *d* or *f* orbitals. For a long time, the simulation of these materials has been considered to be a grand challenge for the Kohn-Sham density functional theory (DFT) [2], the workhorse of the material study, and there is a prevalent belief that DFT is a mean-field theory incapable of the correlated systems. One canonical example is the DFT's inability to predict the insulating behavior of the Mott insulator [3] NiO using the local-density approximation (LDA) [4]. Subsequently, introducing a Hubbard $U$ correction onto the Ni-3$d$ orbitals largely resolves the problem. This fact strengthens a naive belief that electronic correlation is beyond the scope of the DFT approaches.

However, recent advancements show that the DFT approaches can simulate stronger and stronger correlations.

Most noticeably, combining the SCAN (strongly-constrained and appropriately normed) functional [5] with the spin-symmetry-breaking (SSB) technique [6], but without the empirical Hubbard $U$ parameter, we have reliably reproduced many Mott-related characteristics of the prototypical correlated materials, such as FeO and cuprates [7,8,9]. For example, FeO opens a bandgap not only in the antiferromagnetic (AFM) model, which has a long-range magnetic order but also in the disordered-local-moment (DLM) model, which does not rely on magnetic ordering [7]. Similar predictions have been made in the un-doped cuprates like $LaCu_2O_4$ [8] and hole-doped $YBa_2Cu_3O_7$ [9], a group of well-known correlated materials. Moreover, Perdew et al. [6] recently pointed out that "*the spin-symmetry-breaking can reveal strong correlations among the electrons that are present in a symmetry-unbroken wavefunction*" if the broken symmetry persists for a long time. This is usually true for the conventional Mott insulators. For example, according to Philip Anderson [10], the spin-flip span in the AFM solid NiO is typically three years, much longer than the duration of any experimental measurement (e.g., neutron scattering) [6]. These arguments may lay the foundation that the broken symmetry "*may correspond to an actual state*" in condensed matter, as pointed out by Martin, Reining, and Ceperley [11].

Besides the antiferromagnets with spatially *extended* geometries (such as NiO and FeO), a particular class of solids contains *isolated* motifs formed by transition metal-metal bonds [12,13,14]. For example, two and three V atoms form relatively isolated dimer and trimer motifs in $LiVO_2$ [15] and $VO_2$ [16], respectively. In the isolated motifs, the inter-site AFM coupling is quite strong, and the spin interactions are maximally entangled among the sites, which leads to the spin-singlet bonding state. The metal-metal interaction in some solids could be so strong that the transition-metal atoms form "*metallic clusters*" and "*molecules in solids*" [12,13,14], leading to a bond length comparable with or even shorter than that of the elemental metals. The shortest V-V bonds are 2.62 Å in $VO_2$ [16] and 2.56 Å in $LiVO_2$ [17], compared with 2.629 Å [18] in the vanadium metal.

The metal-metal singlet bonding state is characterized by intense spin fluctuations and is observed to be non-magnetic (also referred to as paramagnetic) in experiments. For the application of Kohn-Sham DFT, one can choose between spin-restricted or SSB simulation approaches. The spin-restricted method may be favored by some due to its superior alignment with the non-magnetic reality. However, others might argue that the transition-metal 3$d$ orbitals inherently localize to form a local magnetic moment, suggesting the SSB method, as underscored by Perdew et al., could be more "revealing" [6]. The suitability of these methods for the spin-singlet state remains ambiguous at this juncture. Adding to the complexity is the fact that the robust spin interactions of the singlet bonding might render both the spin-restricted and SSB methods ineffective. This is supported by the recent findings of Streltsov and Khomskii, who noted that "*strong intersite coupling may invalidate the standard single-site starting point for considering magnetism*" [19].

In this work, we select $VO_2$ as an example to evaluate the applicability of Kohn-Sham DFT approaches, including various exchange-correlation functionals on the first four rungs of Jacob's ladder of DFT [20]. $VO_2$ is a fascinating material having three simultaneous transitions around room temperature [21], i.e., (1) the structural transition from the dimerized monoclinic (referred to as $M_1$) phase to the high-temperature rutile (referred to as R) phase, where the dimers break apart, (2) the associated transition of electric transport behaviors from an insulator to a metal, (3) the magnetic transition from a non-magnetic $M_1$ phase to a state with local magnetic moments in R. The concurrent simulation of structural, electrical, and magnetic properties is of utmost importance. Yet, prior DFT research has faced considerable challenges and ambiguity in choosing suitable structural and magnetic models, exchange-correlation functionals, and Hubbard $U$ parameters [21]. There have been instances where researchers confidently predict specific properties while overlooking conflicting outcomes. A particularly perplexing issue pertains to reconciling two ostensibly contradictory properties observed in the $M_1$ $VO_2$: the lack of net magnetic moments and the tendency to develop local magnetism. Addressing this paradox is vital for understanding the driving force of insulator-metal transition, which is still debatable [21] among the Peierls, Mott, or Peierls-Mott-collaborative mechanisms starting from the early days by Goodenough [22].

## 2. Computational details

Most simulations are carried out using the r$^2$SCAN functional [23], which is a revised form of the original SCAN meta-GGA [5], as implemented in the VASP code [24,25]. Both r$^2$SCAN and SCAN are general-purpose functionals on the third rung of Jacob's ladder, yet the former improves the numerical stability significantly. For comparison, we also use other density functionals, including the first rung LDA [4], the second rung generalized-gradient-approximation (GGA) in the form of Perdew-Burke-Ernzerhof (PBE) [26], and the range-separated hybrid functional of Heyd-Scuseria-Ernzerhof (HSE06) [27,28] on the fourth rung. The Heisenberg-type spin exchange parameters, $J$, are extracted using the LKAG method implemented in the TB2J code [29,30]. The chemical bonding property is quantitatively analyzed via the *projected Crystal-Orbital-Hamilton-Population* (pCOHP) [31,32] as implemented in the Lobster code [33]. We find that the effects of spin-orbit coupling play a minor role in the studied properties, and we exclude them from this work. By default, the crystal structures are completely relaxed unless specified otherwise.

## 3. Results and Discussions

### 3.1. Insulating band structure and V-V dimer length

Given that the experimentally observed dimerized M$_1$ phase is a non-magnetic insulator, a key criterion for validating a DFT simulation is the successful reproduction of a gapped band structure, yet without any macroscopic magnetism. This requirement is met by the three theoretical models illustrated in Figure 1. Figure 1(a) represents the non-magnetic (NM) model, where spin-up and spin-down electrons are uniformly balanced across the entire structure. The band structure develops a gap of 0.277 eV [Figure 1(e)] according to the r$^2$SCAN calculation, although it is underestimated compared with an experimentally measured optical bandgap of about 0.6 eV [16]. In the NM model, the emergence of a bandgap underscores the critical role of V-V dimerization. Two V-3$d$ $d_{x^2-y^2}$ electrons strongly hybridize at the center of two V atomic sites, resulting in a dimerization distortion, a critical factor in the Peierls gapping mechanism [34]. The bonding characteristics will be visualized in Section 3.3.

Restricting the spin polarization could be an aggressive hypothesis since the V-3$d$ electrons are intrinsically localized. A circumvention is to allow spin polarization and simultaneously assign an AFM spin configuration to nullify the overall magnetism. This technique is known as SSB for DFT [6,11], and the model is shown in Figure 1(b). The calculated local magnetic moment is 0.96 $\mu_B$, and the bandgap is 0.287 eV [Figure 1(f)]. The gapping is due to the well-known Mott mechanism, which states that double electron occupation of a single atomic site (and thus inter-site electron hopping) is prevented due to significant on-site Coulombic repulsion [3]. Moreover, SSB leads to considerable energy lowering, as established previously [6]. Here, the AFM model is more stable than the NM model by 112.2 meV/formula [Figure 1(d)].

The long-range magnetic ordering in the AFM model, absent in experiments, can be further removed using the *disordered-local-moment* (DLM) model Figure 1(c). Note that the DLM superstructure keeps the AFM pattern within each V-V dimer, which is favored energetically to be discussed in Figure 3, but introduces a disordering pattern between the dimers. The DLM model, without a long-range magnetic ordering, is energetically almost degenerate with the AFM model and has a similar bandgap [Figure 1(d,g)].

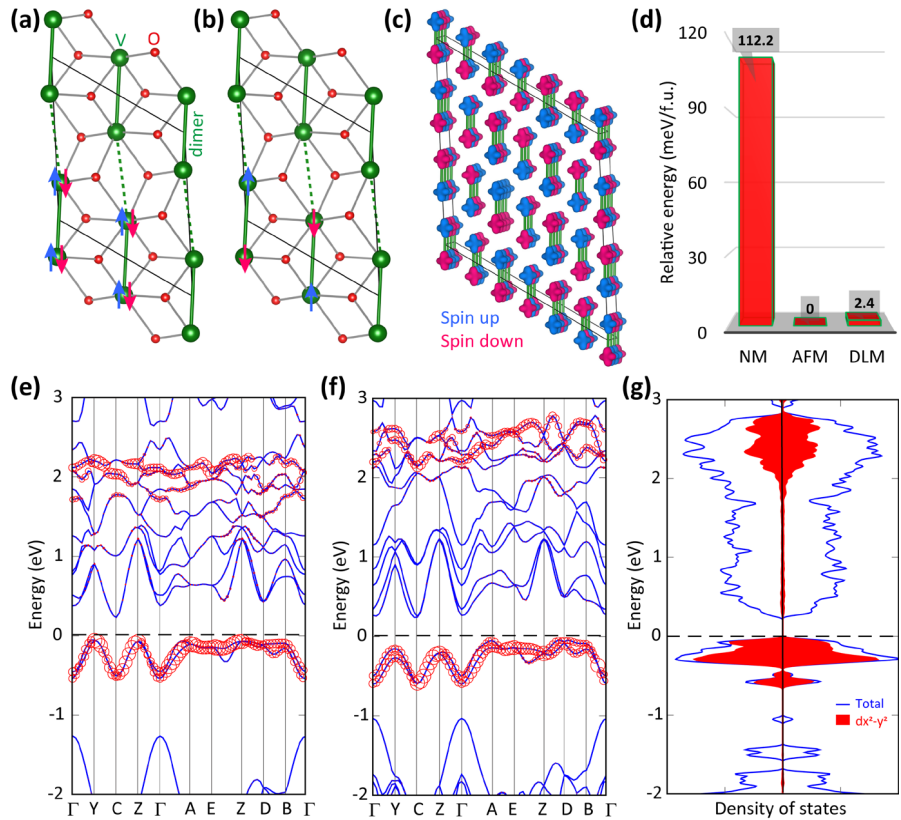

**Figure 1**. Three theoretical models yielding insulating states of $M_1$ $VO_2$ without producing macroscopic magnetism. (a) Spin-restricted non-magnetic (NM) state. The V-V intra- and inter-dimer bonds are denoted as solid and dashed green lines. (b) Each V site is spin-polarized and exhibits antiferromagnetic (AFM) coupling with neighboring sites. (c) Antiferromagnetic coupling within V-V dimers, but the magnetic coupling between different dimers is disordered (i.e., *disordered-local-moment*, DLM). (d) Relative energies of three models. (e,f,g) Corresponding band structures or density-of-states, with the characteristics of $d_{x^2-y^2}$ orbitals highlighted in red. The dashed lines indicate the valence band maximum.

Since both the spin-restricted and SSB methods yield bandgaps, further evaluations, such as those involving crystal structures, become necessary. As shown in Figure 2(a), predicting the V-V dimer length is a challenge task, contrasting with the relatively straightforward prediction of insulating band structures. We conduct comparative studies using four functionals: LDA, PBE, r$^2$SCAN, and HSE06. Besides the spin-restricted NM state, we consider SSB states in two forms: AFM and ferromagnetic (FM) spin configurations. Interestingly, all predicted dimer lengths diverge noticeably from the experimental value of 2.62 Å [16], being either too short or excessively long. An analysis of the magnetic moment [Figure 2(b)] discloses a clear connection: the dimer length is markedly underestimated when the V's magnetic moment is absent, but overestimated once the magnetic moment is stabilized. For instance, in all calculations, LDA cannot stabilize local magnetic moments for V atoms and consistently underestimates the dimer length. PBE cannot stabilize the SSB-AFM model, which eventually converges to the NM state. PBE results in extremely short dimer lengths in the NM state, whereas in the SSB-FM model, which stabilizes local magnetic moments for V atoms, the dimer lengths generated by PBE are too long. The behaviors of r$^2$SCAN and HSE06 are qualitatively alike: the dimer length is even shorter in the NM calculations where local magnetic moments are absent, but excessively long when the moments are stabilized.

Previous studies have established that the SCAN and r$^2$SCAN functionals are generally reliable in describing the geometries of numerous materials [35,36], and noticeable deviations arise in correlated materials due to the persistence of the self-interaction error [37,38]. To mitigate this error in $VO_2$, we combine the $r^2SCAN$ functional with the on-site Hubbard $U$ correction ($r^2SCAN+U$) to reevaluate the structural and magnetic properties using the SSB-AFM model. Interestingly, the V-V intra-dimer length increases with higher values of $U$ [Figure 2(d)], while the distance between dimers decreases accordingly. The dimer structure is wholly disrupted when $U$ reaches 2.0 eV, collapsing into the R phase with equally distributed V ions. Furthermore, a nearly linear correlation exists between the dimer length and the local magnetic moment [Figure 2(e)], indicating that the magnetic moment destabilizes the dimer structure.

Another unexpected result is the stability of the FM spin configuration calculated by the PBE and $r^2SCAN$ functionals, which disagrees with experiments [Figure 2(c)]. The reason will be explained later.

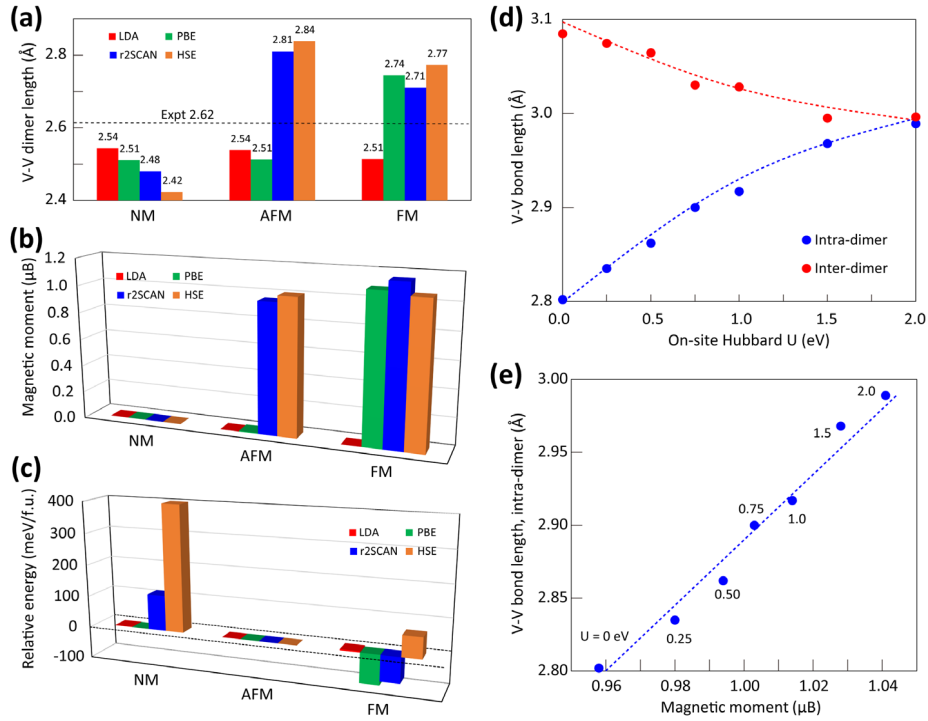

**Figure 2**. Structural, energetic, and magnetic properties of $M_1$ $VO_2$. (a) V-V intra-dimer bond length. Simulations are done with LDA, PBE-GGA, $r^2SCAN$ meta-GGA, and HSE06 hybrid functionals, in combination with spin-restricted NM and the SSB-AFM/FM. (b) Local magnetic moments of V ions. (c) Relative energies of the NM/FM models with respect to the AFM model. (d) V-V bond length from the $r^2SCAN+U$ simulations as a function of Hubbard $U$. (e) V-V intra-dimer length plotted against the local magnetic moments of V ions. Note that the subplots (d,e) are based on the SSB-AFM model.

The preceding evaluations make it clear that spin-restricted calculations utilizing diverse density functionals tend to consistently undervalue the dimer length, which might be attributed to the missing strong correlation. The SSB method is capable of capturing a substantial part of these correlation effects [6], leading to a notable reduction in energy compared to the NM state. However, this SSB method brings forth two new complications in $M_1$ $VO_2$: it destabilizes the dimer structure due to the emergence of local magnetic moments, and (except in HSE) it indicates the FM spin configuration as the most stable. Additionally, it is crucial to note that applying on-site Hubbard $U$ to the SSB model exacerbates the dimer length problem. These findings hint that the issues may extend beyond the tested functionals. The following sections examine a potential factor contributing to these difficulties.

## 3.2. Instability of Néel-ordered state against valence-bond state

In the framework of the SSB-AFM model utilized for DFT simulations, each spin is situated on a discrete atomic site, with their interactions taken into account subsequently. This way of treating spin interactions aligns with the principles of the Mott insulating mechanism, which emphasizes the electronic repulsion among the spin sites. The magnetic ground state is recognized as the Néel state for traditional three-dimensional solids such as NiO and FeO. However, Sachdev et al. [39,40] have established that the *Néel-ordered* state is not the ground state for low-dimensional antiferromagnets due to its instability, which prompts a transition into a *valence-bond state* that highlights intense spin interactions within the low-dimensional motifs. In the case of $M_1$ $VO_2$, the V-V dimers structurally resemble zero-dimensional "molecules". These earlier investigations on the valence-bond state, employing the effective Hamiltonian method, can provide valuable insights into understanding the challenges faced by DFT, assuming a direct relationship exists between the zero-dimensional characteristic of the V-V dimers and the spin interactions within $VO_2$.

We substantiate the above arguments by extracting the spin interaction parameters in $VO_2$, as shown in Figure 3. The central $V_0$ atom is surrounded by ten neighboring atoms, where the $V_1$ and $V_2$ atoms are in line along the dimer chain (the $y$ direction), and the remaining eight are situated off the chain. We first focus on the $J$ parameters for the experimental crystal structure. Notably, the intra-dimer exchange interaction is remarkably strong ($J_{V_0-V_1} = -49.2$ meV), making the other interactions ($J_{V_0-V_2}$ and $J_{V_0-V_x}$ with $x \geq 3$) essentially negligible. Therefore, the strong intra-dimer AFM spin interactions define the zero-dimensional structural motifs, i.e., the V-V dimers. Structural relaxation leads to a significant reduction of the intra-dimer spin interaction to a value of $J_{V_0-V_1} = -15.9$ meV. This reduction aligns with the observed increase in the dimer length [Figure 2(a)]. As the structure relaxes further, the intra-dimer interaction $J_{V_0-V_1}$ begins to converge with the inter-dimer interaction $J_{V_0-V_2}$. When V-V dimerization entirely disappears in the limit, the magnetic ions form a one-dimensional spin chain.

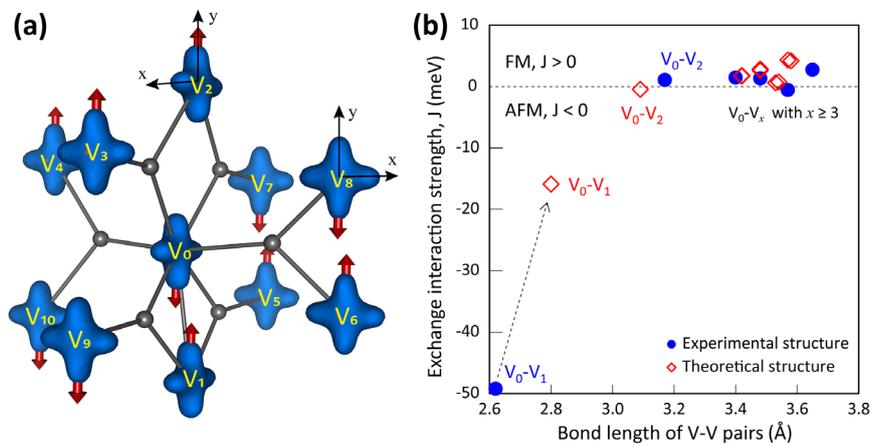

**Figure 3**. Spin interactions in the $M_1$ $VO_2$. (a) Crystal structures showing the central $V_0$ atom and its ten neighbors. The red arrows denote the assumed AFM spin configuration in the simulation. The blue isosurfaces are the electron density of the polarized $d_{x^2-y^2}$ orbitals [41], whose local coordinates are defined by black arrows. (b) Spin exchange parameters extracted from the SSB-AFM model. Two geometries, i.e., the experimental and $r^2$SCAN-relaxed crystal structures, are used in calculations.

The comparative stability of the Néel-ordered and valence-bond states can be illustrated using a one-dimensional illustrative model, as depicted schematically in Figure 4, reproduced from Reference [1]. The instability of one-dimensional spin chains is referred to as the Spin-Peierls problem, which suggests that equally distributed AFM spins within a one-dimensional chain are unstable against spin pairing. The uniformly distributed spins, i.e., the Néel-ordered state, are characterized by an exchange energy:

$$E^{\text{Néel-ordered}} = NJ\langle \mathbf{S}_i \cdot \mathbf{S}_j \rangle = NJ\langle S_i^z S_j^z \rangle = NJ(-\tfrac{1}{4}) = -\tfrac{1}{4}NJ. \quad (1)$$

Here, $N$ represents the number of magnetic interactions, and $J$ represents the exchange strength. Only the $z$-component of spin, $S^z$, contributes to this energy. Contrastingly, spin pairing leads to a *valence-bond state*, and the associated spin interaction energy is:

$$E^{\text{Valence-bond}} = \tfrac{N}{2}J'\langle \mathbf{S}_i \cdot \mathbf{S}_j \rangle = \tfrac{N}{2}J'\langle S_i^x S_j^x + S_i^y S_j^y + S_i^z S_j^z \rangle = \tfrac{N}{2}J'(-\tfrac{3}{4}) = -\tfrac{3}{8}NJ'. \quad (2)$$

In this case, the inter-dimer exchange becomes negligible, causing the valence-bond state to lose half of the spin interactions. However, the valence-bond state could be more stable, because the spin is isotropic and all three spin components contribute equally. It requires that $J' > \tfrac{2}{3}J$, which is generally true as the exchange interaction strengthens after dimerization [see the example of VO$_2$ in Figure 3(b)].

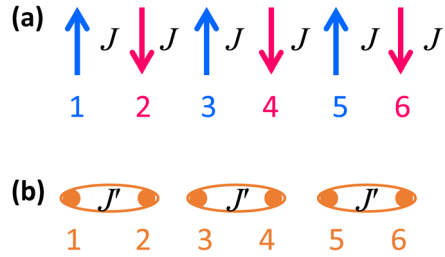

**Figure 4**. Spin-Peierls transition in the one-dimensional spin chain with AFM coupling, reproduced from [1]. (a) Néel-ordered state with uniformly distributed spins. The inter-site exchange parameter is $J$. (b) Valence-bond state with dimer structures. The exchange strength within a dimer is $J'$, while the inter-dimer magnetic interaction is neglectable.

The emergence of the valence-bond state highlights the significance of spin entanglement, a key physical principle intrinsic to many quantum materials [42]. For example, TlCuCl$_3$ [43] and RuCl$_3$ [44] exhibit a quantum phase transition from the Néel-ordered phase to a quantum-disordered phase under external pressure. This disordered phase is predominantly characterized by the valence-bond state. Turning our attention back to M$_1$ VO$_2$, each V$^{4+}$ ion in the V-V dimer motif carries one valence electron, causing the motif to behave akin to a two-electron system. It would be more appropriate to consider VO$_2$ as a *valence-bond solid* because all spins are allocated to localized dimers [1]. This situation contrasts with a resonant-valence-bond (RVB) state, which features dynamically resonating spin interactions [45].

### 3.3. Description of two-electron singlet bonding by Kohn-Sham DFT

Is there any reason preventing the DFT from accurately describing the valence-bond state in M$_1$ VO$_2$? Since Kohn-Sham DFT is formulated in terms of the Kohn-Sham molecular orbitals, it is informative to express the two-electron wave function. The molecular-orbital theory (MOT), pioneered by Mulliken and Hund in the late 1920s [46,47,48,49,50], defines the spin-singlet state of a valence bond in the independent-electron (IE) representation as [51]:

$$\Psi_{\text{MOT,IE}}^{Singlet} = \tfrac{1}{2\sqrt{1+S_{1,2}}}[\varphi_1(\mathbf{r}_1)\varphi_2(\mathbf{r}_2) + \varphi_2(\mathbf{r}_1)\varphi_1(\mathbf{r}_2)][\alpha_1\beta_2 - \beta_1\alpha_2]$$

$$+ \tfrac{1}{2\sqrt{1+S_{1,2}}}[\varphi_1(\mathbf{r}_1)\varphi_1(\mathbf{r}_2) + \varphi_2(\mathbf{r}_1)\varphi_2(\mathbf{r}_2)][\alpha_1\beta_2 - \beta_1\alpha_2] \ . \quad (3)$$

Here, $\varphi(r)$ represents the one-electron atomic wave function, and the double overlap integral is denoted as $S_{1,2} = \langle \varphi_1(r_1)\varphi_2(r_2)|\varphi_1(r_2)\varphi_2(r_1)\rangle$. $\alpha$ and $\beta$ are the spin functions. However, a potential issue may arise in $\Psi_{\text{MOT,IE}}^{Singlet}$. The two ionic terms $\varphi_1(r_1)\varphi_1(r_2) + \varphi_2(r_1)\varphi_2(r_2)$, representing a single atomic site occupied by two electrons, appears at an equally high probability as the other terms. This situation can result in unphysical Coulombic repulsion. While this repulsion is less critical for delocalized electrons (as observed in the H$_2$ molecule with an equilibrium geometry), it becomes particularly prominent in the presence of localized electrons, such as V-3$d$ electrons in VO$_2$.

Given these insights, we can explain the discrepancy in calculating the V-V dimer length between the spin-restricted and SSB models. The V-V bond length is determined by a delicate balance between two interactions. On one side, there is an attractive covalent bonding between neighboring V $d_{x^2-y^2}$ orbitals in a spin-singlet state. This bonding promotes electron overlap around the bond center, resulting in a shortened V-V bond. Conversely, the localization of V-3$d$ orbitals favors spatial separation of the individual V atoms, leading to charge depletion at the bond center and a consequent elongation of the bond length. In the spin-restricted calculation, the suppression of electron localization results in a bias towards covalence, favoring a bond length shorter than the experimental value. Alternatively, the spin polarization of V-3$d$ orbitals in the SSB model better captures the electronic correlation. However, when magnetism is treated using DFT methods, this can overestimate ionic characteristic interactions within a V-V dimer. As a result, this overestimation contributes to a longer bond length than the experimental data.

Our arguments are corroborated by examining the chemical bonding properties using the pCOHP method, specifically through analyzing the negative values (-pCOHP) [31,32]. Figure 5(a) visualizes the bonding interactions between two V-V atoms in a dimer, where all electronic states and the $d_{x^2-y^2}$ orbital interactions are represented by red shades and green lines, respectively. The bonding interaction is principally driven by the valence electrons originating from the $d_{x^2-y^2}$ orbitals, and this interaction is more pronounced in the NM model than in the SSB-AFM model. A comparison of the electron densities of the two models reveals that introducing spin polarization results in electron depletion at the center of the V-V dimer [Figure 5(c)]. Furthermore, the bandwidth of the $d_{x^2-y^2}$ orbital is broader in the spin-restricted calculation (0.54 eV) than in the SSB-AFM case (0.46 eV), with both calculations based on the same experimental crystal structure. The structural effects are demonstrated in Figure 5(b). Structural relaxation tends to enhance bonding in the NM model, while it has the inverse impact on the SSB-AFM model.

The ionic term present in the spin-singlet state can provide insights into the false stability of the FM spin configuration, as observed in the PBE and r$^2$SCAN simulations [Figure 2(c)]. The FM configuration also represents a broken spin symmetry, being one of the three spin-triplet states described by:

$$\Psi_{\text{MOT,IE}}^{Triplet} = \frac{1}{2\sqrt{1-S_{1,2}}}[\varphi_2(r_1)\varphi_1(r_2) - \varphi_1(r_1)\varphi_2(r_2)]\chi_{sym} , \qquad (4)$$

where $\chi_{sym}$ is the triplet symmetric spin function. It is important to note that $\Psi_{\text{MOT,IE}}^{Triplet}$ does not incorporate any additional contribution from the ionic states, unlike $\Psi_{\text{MOT,IE}}^{Singlet}$. Therefore, when calculating the expectation value of a V-V dimer using $\Psi_{\text{MOT,IE}}^{Singlet}$ and $\Psi_{\text{MOT,IE}}^{Triplet}$ wave functions, the latter's energy is likely lower [51].

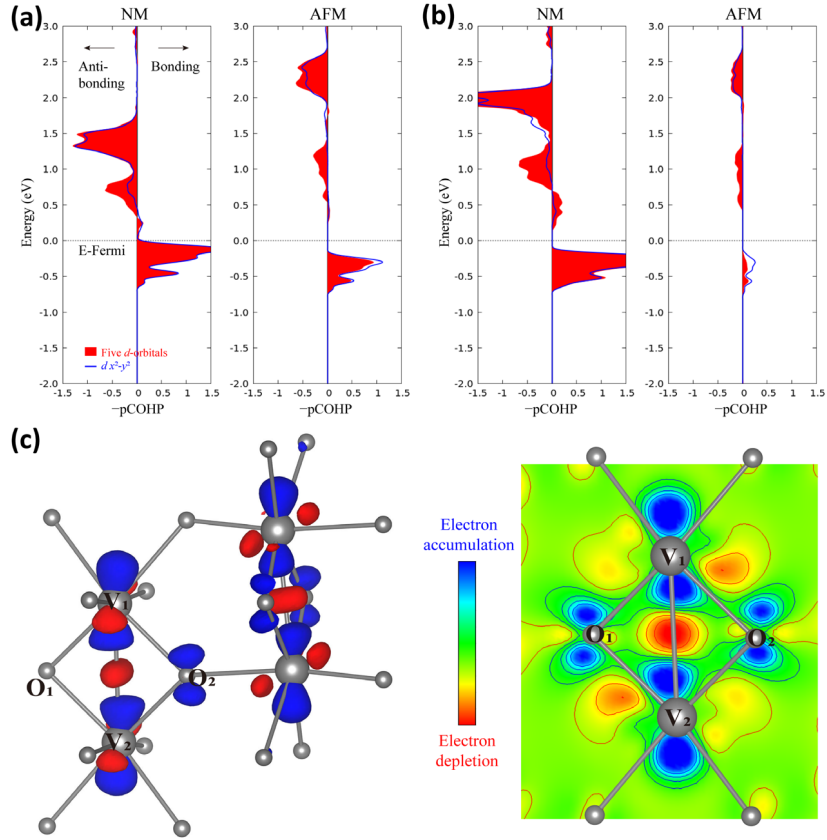

**Figure 5.** Chemical bonding properties of the spin-restricted NM and SSB-AFM models. (a) Diagrams of -pCOHP, showing bonding interactions between two V-V atoms within a dimer. The calculation is done with the experimental crystal structure. (b) The same as subplot (a), except for optimized geometries. (c) Electron density difference of $\Delta n = n^{\mathrm{AFM}} - n^{\mathrm{NM}}$. The experimental crystal structure is used for both states. Note the electron depletion at the V-V dimer center. The right panel is the 2D plot of the $\Delta n$.

For a comprehensive understanding, it's worth mentioning another theory that was developed early on to address spin interactions in chemical bonds. This theory, known as the valence-bond theory (VBT), was developed by Heitler, London, Pauling, and Slater during the 1916-1920s [52,53,54]. The VBT and MOT were considered "rival" theories, but many theoreticians now agree that they complement each other [55,56,57]. In the Heitler-London (HL) limit, the VBT defines the singlet state wave function as [51]:

$$\Psi_{\mathrm{VBT,HL}}^{Singlet} = \frac{1}{2\sqrt{1+S_{1,2}}}[\varphi_1(\mathbf{r}_1)\varphi_2(\mathbf{r}_2) + \varphi_2(\mathbf{r}_1)\varphi_1(\mathbf{r}_2)][\alpha_1\beta_2 - \beta_1\alpha_2] \ . \tag{5}$$

It's important to note the absence of the ionic term in the HL representation. The HL treatment emphasizes the overlap of atomic orbitals in forming the valence-bond state, which increases the likelihood of electrons residing in the bond center. There's a widely accepted belief that the IE and HL wave functions are approximate representations, and the actual electronic state is likely between these extremes. Specifically, $\Psi_{\mathrm{VB,HL}}^{Singlet}$ is thought to be more appropriate for describing the equilibrium $H_2$ molecule [51]. In the case of $M_1$ $VO_2$, cluster dynamical mean-field theory, in conjunction with the DFT scheme (DFT+cDMFT) [58,59,60], suggested that the V-V dimer is close to the HL limit [59]. However, we propose that the inherently localized nature of the 3$d$ electrons may somewhat invalidate the HL representation for $M_1$ $VO_2$.

### 3.4. Spin-symmetry-breaking in the intermediate $M_2$ VO$_2$: An imperfect yet valuable approach

Finally, we present a situation in which SSB is highly advantageous, even though imperfect. In addition to the well-known $M_1$ and R phases, VO$_2$ can also form an intermediate phase under pressure or chemical doping conditions. Half of the V atoms form dimer chains in this phase, while the remaining V atoms are evenly dispersed. Experimentally, $M_2$ VO$_2$ is an insulator [21].

Figure 6 demonstrates the band structures calculated using the spin-restricted NM and SSB-AFM models, producing a metallic and an insulating phase, respectively. By projecting the orbital characters onto the band structure, we can distinguish the varying behaviors of the dimerized and non-dimerized V atoms. In the NM model, the dimerized V atoms do not contribute electrons at the Fermi level, while the non-dimerized V atoms account for the metallic nature. This metallic character arises due to the suppression of the Mott gapping mechanism as a result of the spin restriction. Conversely, in the SSB model, both the dimerized and non-dimerized V atoms can establish gaps via the Mott mechanism. It is important to acknowledge that, based on the discussion in previous sections, these two types of V atoms should ideally rely on different gapping mechanisms. Therefore, within the Kohn-Sham DFT framework, SSB is a valuable method for producing an insulating band structure of $M_2$ VO$_2$.

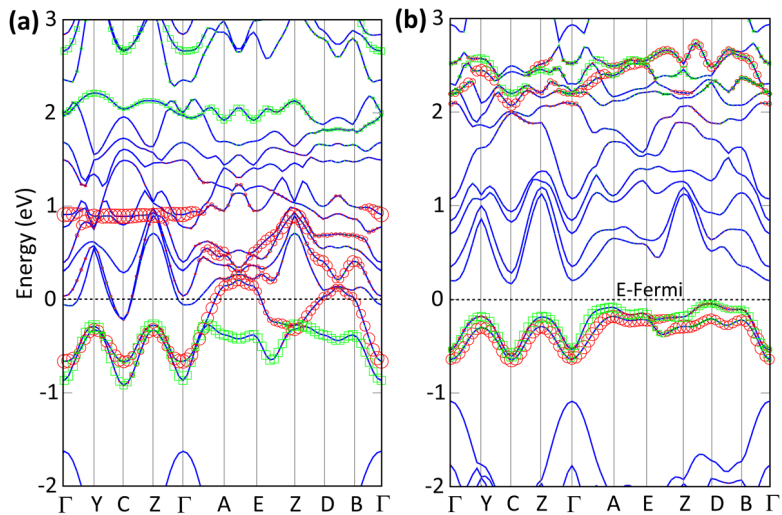

**Figure 6.** Band structure of $M_2$ VO$_2$. (a) Spin-restricted NM model. (b) SSB-AFM model. The red circles (green squares) represent V $d_{x^2-y^2}$ orbitals of the non-dimerized (dimerized) chain.

## 4. Summary

This study addresses a fundamental difficulty in characterizing the dimer structure of $M_1$ VO$_2$, primarily utilizing the advanced r$^2$SCAN density functional. The metal-metal singlet state is marked by robust covalent bonding and intense spin fluctuations within a V-V dimer. However, there is a high degree of localization of V-3$d$ electrons around equilibrium atomic positions in VO$_2$, which is unlike the H$_2$ molecule in a stretched condition.

These aspects present a significant challenge to Kohn-Sham DFT methods. On one hand, the spin-restricted model severely inhibits the localization of the V-3$d$ electron, which results in a high energy. The predicted dimer length is strongly underestimated. On the other hand, the SSB model respects the electron localization and thus provides a more accurate description of the electronic correlation energy. However, the predicted dimer length tends to be too long due to the excessively strong electronic repulsions, as seen in the ionic term when presenting the two-electron wave function within the molecular-orbital theory framework. The spin-restricted and SSB models represent two extreme scenarios, with the actual V-V bonding likely existing somewhere in between.

For the ground-state total energy, the inadequacies of the Kohn-Sham determinant as an approximation to the true wavefunction should (except in special cases, including failure of the exact density to be non-interacting v-representable) be exactly compensated by the sum of the Hartree energy and the exact density functional for the exchange-correlation energy. But we only have approximate functionals, which compensate imperfectly. R$^2$SCAN was constructed to respect most known mathematical properties of the exact functional, but none of the functionals tested here is exact for all one-electron densities. Thus, VO$_2$ stands as a challenge to any self-interaction correction to r$^2$SCAN that exists (e.g., the $+U$ correction [61] or the full Fermi-Löwdin self-interaction correction [62]) or will be developed.

These findings highlight the continuing challenges faced by the DFT approaches in accurately describing the metal-metal singlet bonding. A reliable treatment should allow for breaking the spin-symmetry while capturing intense spin interactions more accurately. Currently, the combination of r$^2$SCAN with SSB is an imperfect yet valuable approach, as demonstrated in the intermediate M$_2$ VO$_2$ phase. Our results on VO$_2$ also have broader implications for calculations of quantum antiferromagnets in general, such as quantum spin liquid systems.

## Acknowledgments

We thank Youjin Deng, Peihong Zhang, and Yuanyao He for their fruitful discussions. YZ is supported by the Fujian Natural Science Foundation (2023J02032), Guangdong Natural Science Foundation (2021A1515010049), and the National Natural Science Foundation of China (11904156). The work of JPP was supported by the U.S. National Science Foundation under Grant No. DMR-1939528, and by the U.S. Department of Energy. Basic Energy Sciences, under Grant. No. DE-SC18331. J.S. acknowledges the support of the U.S. Office of Naval Research (ONR) Grant No. N00014-22-1-2673.

## References

[1] Daniel Khomskii, *Transition metal compounds*. 2014: Cambridge University Press.
[2] W. Kohn, L. J. Sham, *Self-Consistent Equations Including Exchange and Correlation Effects.* Phys. Rev. 140, A1133 (1965).
[3] Nevill Mott, *Metal-insulator transitions*. 2004: CRC Press.
[4] P. Hohenberg, W. Kohn, *Inhomogeneous Electron Gas.* Phys. Rev. 136, B864 (1964).
[5] Jianwei Sun, Adrienn Ruzsinszky, John P Perdew, *Strongly Constrained and Appropriately Normed Semilocal Density Functional.* Phys. Rev. Lett. 115, 036402 (2015).
[6] John P. Perdew, Adrienn Ruzsinszky, Jianwei Sun, Niraj K. Nepal, Aaron D. Kaplan, *Interpretations of ground-state symmetry breaking and strong correlation in wavefunction and density functional theories.* Proc. Natl. Acad. Sci. U.S.A. 118, e2017850118 (2021).
[7] Yubo Zhang, James Furness, Ruiqi Zhang, Zhi Wang, Alex Zunger, Jianwei Sun, *Symmetry-breaking polymorphous descriptions for correlated materials without interelectronic U.* Phys. Rev. B. 102, 045112 (2020).
[8] James W. Furness, Yubo Zhang, Christopher Lane, Ioana Gianina Buda, Bernardo Barbiellini, Robert S. Markiewicz, Arun Bansil, Jianwei Sun, *An accurate first-principles treatment of doping-dependent electronic structure of high-temperature cuprate superconductors.* Communications Physics. 1, 11 (2018).
[9] Yubo Zhang, Christopher Lane, James W. Furness, Bernardo Barbiellini, John P. Perdew, Robert S. Markiewicz, Arun

Bansil, Jianwei Sun, *Competing stripe and magnetic phases in the cuprates from first principles.* Proc. Natl. Acad. Sci. U.S.A. 117, 68 (2020).

[10] P. W. Anderson, *More Is Different.* Science. 177, 393 (1972).

[11] Richard M Martin, Lucia Reining, David M Ceperley, *Interacting electrons.* 2016: Cambridge University Press.

[12] S. V. Streltsov, D. I. Khomskii, *Orbital physics in transition metal compounds: new trends.* Physics-Uspekhi. 60, 1121 (2017).

[13] Daniel I. Khomskii, Sergey V. Streltsov, *Orbital Effects in Solids: Basics, Recent Progress, and Opportunities.* Chem. Rev. 121, 2992 (2021).

[14] D. I. Khomskii, *Review—Orbital Physics: Glorious Past, Bright Future.* ECS J. Solid State Sci. Technol. 11, 054004 (2022).

[15] H. F. Pen, J. van den Brink, D. I. Khomskii, G. A. Sawatzky, *Orbital Ordering in a Two-Dimensional Triangular Lattice.* Phys. Rev. Lett. 78, 1323 (1997).

[16] V. Eyert, *The metal-insulator transitions of VO2: A band theoretical approach.* Ann. Phys. 514, 650 (2002).

[17] Takaaki Jin-no, Yasuhiro Shimizu, Masayuki Itoh, Seiji Niitaka, Hidenori Takagi, *Orbital reformation with vanadium trimerization in $d^2$ triangular lattice $LiVO_2$ revealed by $^{51}V$ NMR.* Phys. Rev. B. 87, 075135 (2013).

[18] E. SÁNdor, W. A. Wooster, *Extra Streaks in the X-ray Diffraction Pattern of Vanadium Single Crystals.* Nature. 182, 1435 (1958).

[19] Sergey V. Streltsov, Daniel I. Khomskii, *Covalent bonds against magnetism in transition metal compounds.* Proc. Natl. Acad. Sci. U.S.A. 113, 10491 (2016).

[20] John P. Perdew, Karla Schmidt, *Jacob's ladder of density functional approximations for the exchange-correlation energy.* AIP Conf. Proc. 577, 1 (2001).

[21] Jean-Paul Pouget, *Basic aspects of the metal–insulator transition in vanadium dioxide VO2: a critical review.* Comptes Rendus. Physique. 22, 37 (2021).

[22] John B. Goodenough, *The two components of the crystallographic transition in VO2.* J. Solid State Chem. 3, 490 (1971).

[23] James W. Furness, Aaron D. Kaplan, Jinliang Ning, John P. Perdew, Jianwei Sun, *Accurate and Numerically Efficient r2SCAN Meta-Generalized Gradient Approximation.* J. Phys. Chem. Lett. 11, 8208 (2020).

[24] G. Kresse, J. Furthmüller, *Efficiency of ab-initio total energy calculations for metals and semiconductors using a plane-wave basis set.* Comput. Mater. Sci. 6, 15 (1996).

[25] G. Kresse, J. Furthmüller, *Efficient iterative schemes for ab initio total-energy calculations using a plane-wave basis set.* Phys. Rev. B. 54, 11169 (1996).

[26] John P. Perdew, Kieron Burke, Matthias Ernzerhof, *Generalized Gradient Approximation Made Simple.* Phys. Rev. Lett. 77, 3865 (1996).

[27] Joachim Paier, Martijn Marsman, K Hummer, Georg Kresse, Iann C Gerber, János G Ángyán, *Screened hybrid density functionals applied to solids.* J. Chem. Phys. 124, 154709 (2006).

[28] Jochen Heyd, Gustavo E Scuseria, Matthias Ernzerhof, *Hybrid functionals based on a screened Coulomb potential.* J. Chem. Phys. 118, 8207 (2003).

[29] A. I. Liechtenstein, M. I. Katsnelson, V. P. Antropov, V. A. Gubanov, *Local spin density functional approach to the theory of exchange interactions in ferromagnetic metals and alloys.* J. Magn. Magn. Mater. 67, 65 (1987).

[30] Xu He, Nicole Helbig, Matthieu J. Verstraete, Eric Bousquet, *TB2J: A python package for computing magnetic interaction parameters.* Comput. Phys. Commun. 264, 107938 (2021).

[31] Richard Dronskowski, Peter E. Bloechl, *Crystal orbital Hamilton populations (COHP): energy-resolved visualization of chemical bonding in solids based on density-functional calculations.* J. Phys. Chem. 97, 8617 (1993).

[32] Volker L. Deringer, Andrei L. Tchougréeff, Richard Dronskowski, *Crystal Orbital Hamilton Population (COHP) Analysis*

*As Projected from Plane-Wave Basis Sets.* The Journal of Physical Chemistry A. 115, 5461 (2011).

[33] Stefan Maintz, Volker L. Deringer, Andrei L. Tchougréeff, Richard Dronskowski, *LOBSTER: A tool to extract chemical bonding from plane-wave based DFT.* J. Comput. Chem. 37, 1030 (2016).

[34] Rudolf Peierls, *More surprises in theoretical physics*. 1991: Princeton University Press.

[35] Jianwei Sun, Richard C. Remsing, Yubo Zhang, Zhaoru Sun, Adrienn Ruzsinszky, Haowei Peng, Zenghui Yang, Arpita Paul, Umesh Waghmare, Xifan Wu, Michael L. Klein, John P. Perdew, *Accurate first-principles structures and energies of diversely bonded systems from an efficient density functional.* Nature Chemistry. 8, 831 (2016).

[36] Yubo Zhang, Jianwei Sun, John P. Perdew, Xifan Wu, *Comparative first-principles studies of prototypical ferroelectric materials by LDA, GGA, and SCAN meta-GGA.* Phys. Rev. B. 96, 035143 (2017).

[37] Yubo Zhang, Daniil A. Kitchaev, Julia Yang, Tina Chen, Stephen T. Dacek, Rafael A. Sarmiento-Pérez, Maguel A. L. Marques, Haowei Peng, Gerbrand Ceder, John P. Perdew, Jianwei Sun, *Efficient first-principles prediction of solid stability: Towards chemical accuracy.* npj Computational Materials. 4, 9 (2018).

[38] Yubo Zhang, James W. Furness, Bing Xiao, Jianwei Sun, *Subtlety of TiO2 phase stability: Reliability of the density functional theory predictions and persistence of the self-interaction error.* J. Chem. Phys. 150, 014105 (2019).

[39] N. Read, Subir Sachdev, *Spin-Peierls, valence-bond solid, and Neel ground states of low-dimensional quantum antiferromagnets.* Phys. Rev. B. 42, 4568 (1990).

[40] N. Read, Subir Sachdev, *Valence-bond and spin-Peierls ground states of low-dimensional quantum antiferromagnets.* Phys. Rev. Lett. 62, 1694 (1989).

[41] A. Liebsch, H. Ishida, G. Bihlmayer, *Coulomb correlations and orbital polarization in the metal-insulator transition of VO2.* Phys. Rev. B. 71, 085109 (2005).

[42] Subir Sachdev, *The quantum phases of matter.* arXiv:1203.4565, (2012).

[43] Ch Rüegg, B. Normand, M. Matsumoto, A. Furrer, D. F. McMorrow, K. W. Krämer, H. U. Güdel, S. N. Gvasaliya, H. Mutka, M. Boehm, *Quantum Magnets under Pressure: Controlling Elementary Excitations in TlCuCl3.* Phys. Rev. Lett. 100, 205701 (2008).

[44] G. Bastien, G. Garbarino, R. Yadav, F. J. Martinez-Casado, R. Beltrán Rodríguez, Q. Stahl, M. Kusch, S. P. Limandri, R. Ray, P. Lampen-Kelley, D. G. Mandrus, S. E. Nagler, M. Roslova, A. Isaeva, T. Doert, L. Hozoi, A. U. B. Wolter, B. Büchner, J. Geck, J. van den Brink, *Pressure-induced dimerization and valence bond crystal formation in the Kitaev-Heisenberg magnet RuCl3.* Phys. Rev. B. 97, 241108 (2018).

[45] P. W. Anderson, *Resonating valence bonds: A new kind of insulator?* Mater. Res. Bull. 8, 153 (1973).

[46] Robert S. Mulliken, *The Assignment of Quantum Numbers for Electrons in Molecules. I.* Phys. Rev. 32, 186 (1928).

[47] Robert S. Mulliken, *The Assignment of Quantum Numbers for Electrons in Molecules. II. Correlation of Molecular and Atomic Electron States.* Phys. Rev. 32, 761 (1928).

[48] Robert S. Mulliken, *The Assignment of Quantum Numbers for Electrons in Molecules. III. Diatomic Hydrides.* Phys. Rev. 33, 730 (1929).

[49] F. Hund, *Zur Deutung der Molekelspektren. IV.* Z. Phys. 51, 759 (1928).

[50] F. Hund, *Zur Frage der chemischen Bindung.* Z. Phys. 73, 1 (1932).

[51] Joachim Stöhr, Hans Christoph Siegmann, *Magnetism - From fundamentals to nanoscale dynamics*. Vol. 5. 2006: Springer, Berlin, Heidelberg. 236.

[52] Gilbert N. Lewis, *THE ATOM AND THE MOLECULE*. J. Am. Chem. Soc. 38, 762 (1916).

[53] Linus Pauling, *The nature of the chemical bond*. 3 ed. 1939: Cornell University Press.

[54] W. Heitler, F. London, *Wechselwirkung neutraler Atome und homöopolare Bindung nach der Quantenmechanik.* Z. Phys. 44, 455 (1927).

[55] John Morrison Galbraith, Sason Shaik, David Danovich, Benoît Braïda, Wei Wu, Philippe Hiberty, David L. Cooper, Peter B. Karadakov, Thom H. Dunning, Jr., *Valence Bond and Molecular Orbital: Two Powerful Theories that Nicely*

*Complement One Another.* J. Chem. Educ. 98, 3617 (2021).

[56] Sason Shaik, David Danovich, Philippe C. Hiberty, *Valence Bond Theory—Its Birth, Struggles with Molecular Orbital Theory, Its Present State and Future Prospects.* Molecules. 26, 1624 (2021).

[57] Gernot Frenking, Sason Shaik, *The chemical bond: fundamental aspects of chemical bonding*. Vol. 1. 2014: John Wiley & Sons.

[58] W. H. Brito, M. C. O. Aguiar, K. Haule, G. Kotliar, *Dynamic electronic correlation effects in NbO2 as compared to VO2.* Phys. Rev. B. 96, 195102 (2017).

[59] S. Biermann, A. Poteryaev, A. I. Lichtenstein, A. Georges, *Dynamical Singlets and Correlation-Assisted Peierls Transition in VO2.* Phys. Rev. Lett. 94, 026404 (2005).

[60] Bence Lazarovits, Kyoo Kim, Kristjan Haule, Gabriel Kotliar, *Effects of strain on the electronic structure of $VO_2$.* Phys. Rev. B. 81, 115117 (2010).

[61] Matteo Cococcioni, Stefano de Gironcoli, *Linear response approach to the calculation of the effective interaction parameters in the LDA+U method.* Phys. Rev. B. 71, 035105 (2005).

[62] Mark R. Pederson, Adrienn Ruzsinszky, John P. Perdew, *Communication: Self-interaction correction with unitary invariance in density functional theory.* J. Chem. Phys. 140, 121103 (2014).